 \documentclass[final,3p,times,twocolumn]{elsarticle}

\usepackage{graphicx}

\usepackage{amssymb}

\usepackage{lineno}

\journal{ChemPhys}

\begin{document}

\begin{frontmatter}
\title{Inertial Brownian motors driven by  biharmonic  signals\tnoteref{birth}}
\author[pl]{Lukasz Machura}
\author[pl]{Marcin Kostur}
\author[pl]{Jerzy \L uczka\corref{cor}}
\ead{jerzy.luczka@us.edu.pl}
\ead[http://fizyka.us.edu.pl]{fizyka.us.edu.pl}
\address[pl]{Institute of Physics, University of Silesia, 40--007 Katowice, Poland}

\cortext[cor]{Corresponding author.}
\tnotetext[birth]{On the occasion of 60th birthday of Prof. Peter H\"anggi}

\begin{abstract}
We study  transport properties of an  inertial Brownian particle  moving in  viscous  symmetric 
periodic structures and driven by an oscillating signal  of two harmonic components. 
We analyze the influence of  symmetric, antisymmetric and asymmetric signals on directed transport 
and reveal the shift symmetry of the stationary averaged velocity of the Brownian particle with 
respect to the relative phase of two components of the signal. The shift symmetry holds true in all regimes. 
\end{abstract}


\end{frontmatter}

\linenumbers

\section{Introduction}

Recent progress  in the highly controlled fabrication of small  structures opens  new prospects 
 for miniaturization of devices, machines, engines, etc.    Processes in such systems can exhibit radically different properties than at the macroscopic level. For example at  the  microscopic scale,  immanently there is a world of fluctuations which cannot be eliminated or even reduced. However, it can be exploited. A good example are biological motors like kinesin or dynein which exploit thermal fluctuations for their directed movement by the ratchet mechanism \cite{bio}.  
At the microscopic or mesoscopic levels, ways and means  of generation and control of particle 
transport  are important issues for both theorists \cite{HanMar2009} and experimentalists
 \cite{Linke2002}.   In literature, there are many 
suggestions and examples how to generate a directed movement of particles \cite{HanMar2009,Luczka1995}.  
Much more difficult problem is related to 
a  precise control of  transport. In the paper, we study an archetype of transport in (spatially)
  periodic systems which is described by a Langevin equation. In this modeling, we know what 
  conditions have to be fulfilled in order to generate a directed motion of a Brownian particle. 
Moreover, properties of this system can be  experimentally verified in a setup consisting 
of a resistively and capacitively shunted Josephson junction device 
 \cite{MacKos2007,MacKos2008,Nagel2008}. 
It is possible because the underlying dynamics can conveniently be described by an equivalent 
equation of motion  in  the  Stewart-McCumber model \cite{stewart,mccumber,junction,kautz}. In our previous papers  \cite{MacKos2007,MacKos2008},  
we have studied  the system driven by a time-periodic force $G(t)$ which is the simplest 
harmonic signal  $G(t)=A \cos (\Omega t)$ 
(or $G(t)=A \sin (\Omega t)$), where $A$  and $\Omega$ are the amplitude and angular 
frequency of the signal, respectively.  We have shown that,  when additionally 
a constant force $F$ is applied, 
 anomalous transport  in  experimentally wide regimes  can be observed: 
absolute negative  mobility  near zero
value of $F$ ( a linear response regime), negative mobility  in the nonlinear 
response regime and negative differential mobility. 
In this paper we extend the analysis  by considering the biharmonic driving.   
However, we assume that the constant force  $F = 0$. 

The paper is organized as follows.  In Sec. 2, we
present the  Langevin equation determining  dynamics of the Brownian
particle in presence of  $\delta$-correlated thermal fluctuations. Next, in Sec. 3,
we address the problem of influence  of   the second harmonics  on transport of the Brownian particle. 
In the parameter space, we reveal   reach  transport behavior.  
 Sec. 4  provides summary and some conclusions.

\section{Langevin dynamics}

We  study the motion of a classical particle of mass $m$
moving in the periodic, symmetric one-dimensional potential $V(x) = \Delta V \sin(2\pi x/L)$
of  the period $L$ and a barrier height $2\Delta V$. 
The particle is driven by an unbiased time-periodic  \emph{biharmonic} force 
\begin{eqnarray}
 \label{biforce}
G(t) = A [ \sin (\Omega t) + \epsilon \sin (2 \Omega t + \phi) ], 
\end{eqnarray}
where $\epsilon$ is the ratio of the second harmonic amplitude to the fundamental amplitude $A$
and the relative phase $\phi$ determines the time symmetry of the system. 
Additionally, the particle is subjected to the thermal noise. 
Dynamics of a such defined   Brownian motor   is  governed by the Langevin
equation for the coordinate $x=x(t)$ of the Brownian particle which has the form 
\cite{hanggi1982}
\begin{eqnarray}
 \label{Lan}
 m \ddot x + \gamma \dot x = -V'(x) + G(t) + \sqrt{2\gamma kT} \; \xi(t),
\end{eqnarray}
where the dot denotes a differentiation with respect  to time and prime denotes a differentiation with respect to the argument of the potential 
 $V(x)$. The parameter $\gamma$ is the friction coefficient, $T$
denotes temperature, and $k$ is the Boltzmann constant. 
Thermal fluctuations are modeled by the
zero-mean  Gaussian white noise $\xi(t)$ with the  correlation function 
$\langle \xi(t)\xi(s)\rangle = \delta(t-s)$.  

We introduce dimensionless variables. The natural length scale is determined
by the period $L$ of the  potential $V(x)$. The dynamics possesses
several time scales.  We define the characteristic time $\tau_0$ 
determined  from the Newton equation, $m\ddot x=-V'(x)$, by inserting 
characteristic quantities, namely, $mL/\tau_0^2 = \Delta V/L$; hence $\tau_0^2
= mL^2/\Delta V$.  The dimensionless variables thus read:
\begin{eqnarray}
\hat{x} = \frac{x}{L}, \qquad \hat{t} = \frac{t} {\tau_0}.
\end{eqnarray}
The dimensionless Langevin dynamics consequently assumes the form
\begin{equation}
\ddot{\hat x} + \hat{\gamma} \dot{\hat x} =- \hat{V}'(\hat{x}) + 
g(\hat{t}) + \sqrt{2\hat{\gamma}D_0} \; \hat{\xi} (\hat{t}),
\label{NLbw}
\end{equation}
where
\begin{itemize}
\item the re-scaled friction coefficient $\hat{\gamma} = (\gamma / m)
\tau_0$ is the ratio of the two characteristic time scales, $\tau_0$
and the relaxation time scale of the velocity degree of freedom, i.e.,
$\tau_L = m/\gamma$,
\item the re-scaled potential 
\begin{eqnarray}
 \label{pot}
 \hat{V}(\hat{x})=V(x)/\Delta V = \sin(2\pi \hat x)
\end{eqnarray}
 assumes the period $1$ and the barrier height  2,
\item the scaled external time-periodic force 
\begin{equation}
g(\hat{t}) = a [ \sin (\omega \hat{t}) + \varepsilon \sin (2 \omega \hat{t} + \phi) ]
\label{gt}
\end{equation}
where the signal has the re-scaled amplitudes $a = A L / \Delta V$ and  
$\varepsilon = \epsilon / \Delta V$ and  
 the dimensionless angular frequencies $\omega = \Omega \tau_0$,
\item the re-scaled, zero-mean Gaussian white noise forces
$\hat{\xi}(\hat{t})$ obey
$\langle\hat{\xi}(\hat{t})\hat{\xi}(\hat{s})\rangle=\delta(\hat{t}-\hat{s})$
with a re-scaled noise intensity $D_0 = kT / \Delta V$.
\end{itemize}
In the following, mostly for the sake of simplicity, we shall  use only 
dimensionless variables and shall omit the ``hat''-notation  in all
quantities.  

Transport properties  in systems driven by this type of external stimulus 
have been theoretically studied mainly in the overdamped regime \cite{Borromeo2005a,Borromeo2005b,Borromeo2006}, for moderate damping \cite{Breymayer1984},
both experimentally and theoretically for cold atoms in the optical lattices
 \cite{Renzoni2005,Renzoni2008,Denisov2010},
and for  driven Josephson junctions \cite{Monaco1990}.

\section{Influence of the second harmonic of the driving} 

From the symmetry considerations it follows that the long-time averaged velocity $v$ of the 
Brownian motor is equal to zero if it is driven only by one  harmonic, i.e. when  
$\varepsilon =0$ in Eq. (\ref{gt}).  In order to generate a directed motion of the 
motor, one has to include the second harmonic. 
Therefore we pose here the question: what is the influence of the second component
($\varepsilon \ne 0$)  of the external  force $g(t)$ 
on  transport properties of the Brownian particle described by  Eq.  (\ref{NLbw}).

Nonlinearity and three-dimensional phase space $ (x,  y=\dot x, z=\omega t) $ make the system (\ref{NLbw}) possible
to behave chaotically in the deterministic case ($ D_0 = 0 $). Many features depend strongly on 
the shape of  basins of attraction. If we however plug the temperature on, it is very likely 
that we destroy the present scene of attractors and release the possibility for the system to 
proceed not only with attractors but more importantly with the deterministic unstable orbits. 
This situation is extremely complicated and can change from point to point in the  five-dimensional
parameter space $ \{\gamma, \omega, a, \varepsilon, D_0\} $. It is almost impossible to find all
 features for  such a system; therefore the goal of this work is focused
only on the generic influence of  the biharmonicity  parameter   $ \varepsilon$. In fact, 
one is able to tangle the picture even more by setting the frequency of the second harmonic 
in $g(t)$ free, but authors feel that this is unnecessarily in this very work.

In the following we will fix the dimensionless temperature to  the value $D_0 = 0.001$ and  
focus on the stochastic (not deterministic)  properties.

\subsection{Numerical experiment}

In order to establish the influence of the second harmonic of the 
driving force on  transport properties we have carried out 
comprehensive numerical simulations. We have employed Stochastic 
Runge--Kutta algorithm of the $ 2^{nd} $ order with the time step of 
$[10^{-3} \div 10^{-4}](2 \pi / \omega)$. All numerical calculations have been 
performed using CUDA environment on desktop computing processor NVIDIA 
Tesla C1060. This gave us a possibility to speed the numerical 
calculations up to few hundreds times more than on typical modern 
CPUs. More details on this very efficient method can be found in the 
work \cite{cuda}.
 
We focus on the asymptotic current or long--time averaged velocity  $v$  of 
the Brownian particle. Averaging was performed over $10^3 - 10^6$ 
different realizations and over one period of the external driving 
force $T = 2 \pi / \omega$. We choose all initial positions and 
velocities to be  uniformly distributed over one potential period  $[0, 1]$ 
and the interval $v \in [-2,2]$,  respectively.

\subsection{Role of symmetry  in  time domain}

\begin{figure}[htbp]
\includegraphics[width=0.9\linewidth]{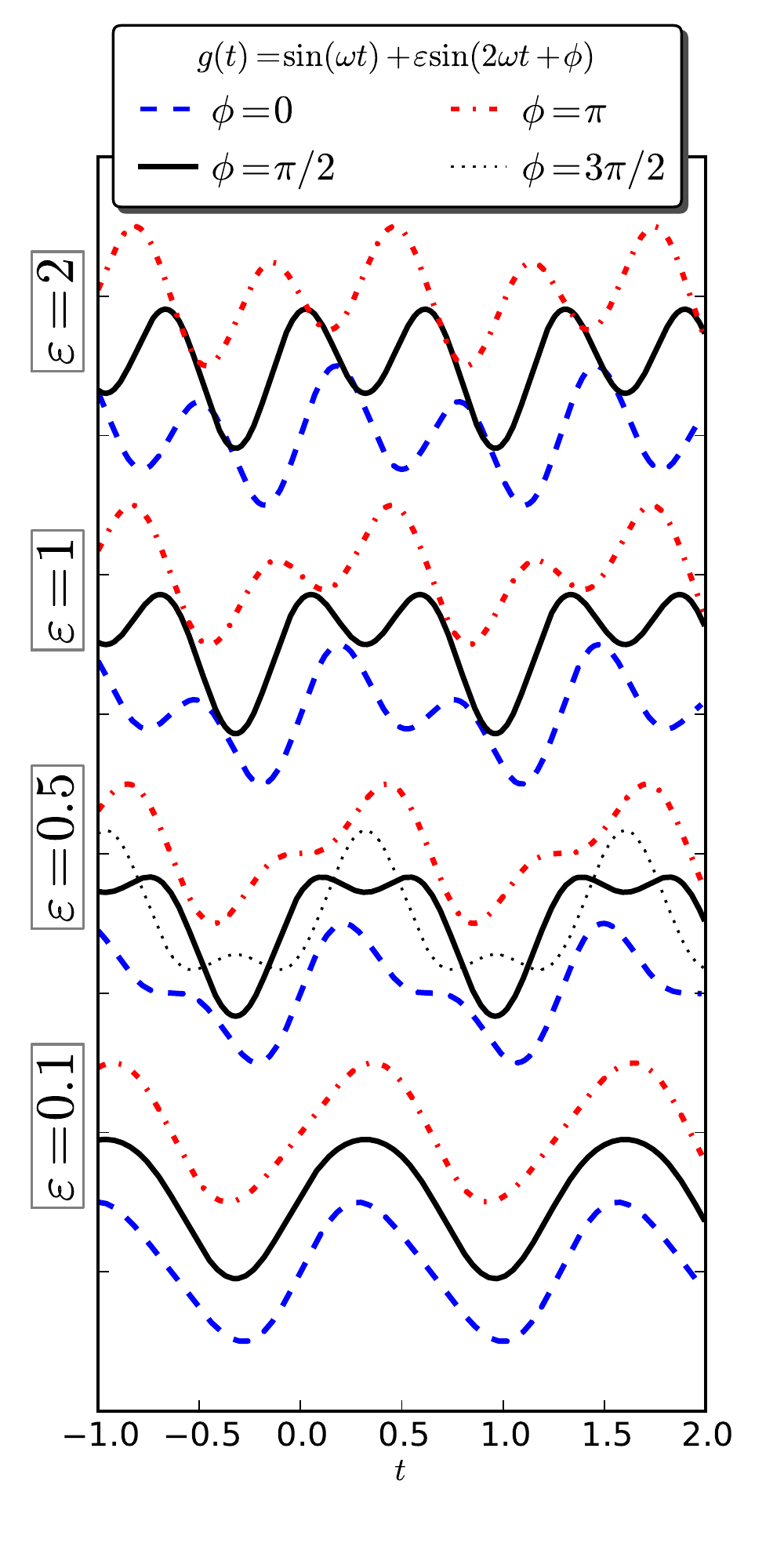} 
\caption{(color online) Dimensionless external  ac  driving
$g(t) = a[\sin(\omega t) + \varepsilon \sin (2\omega t + \phi)]$ for the fundamental amplitude $ a=1$, 
four different relative amplitudes  of the second harmonics: $ \varepsilon = 0.1, 0.5, 1, 2 $
and selected  values of the relative phase:  $ \phi = 0 $ (blue dashed), $\pi/2 $ (black solid), 
$ \pi $ (red dotted-dashed) and $ 3\pi/2 $ (black dotted). 
For arbitrary values of $a$ and $ \varepsilon $, the 
ac  driving possesses the  time reflection symmetry for  $ \phi = \pi /2$ 
and  $ \phi = 3 \pi /2 $.  For $\phi =0$ and $\phi =\pi$  the driving is antisymmetric.  For other  values of the relative phase  $\phi$ the driving is asymmetric.
}
\label{fig0}
\end{figure}

Properties of the time dependent driving force $g(t)$  in Eq. (\ref{gt}) determine whether  
the Brownian particle is transported in the long-time regime, i.e. whether $v = 0$ or $v\ne 0$. 
We can distinguish two special cases   of the force $g(t)$ . \\
(i)  The first case  is when there is  such  $t_0$  that $g(t_0 +t) = g(t_0 - t)$.  It  means that the driving is symmetric or invariant under the time-inversion transformation, see solid and dotted  lines in  
Fig.  \ref{fig0}. 
 \\
(ii)  The second case is when  there is  such $t_1$ that $g(t_1 +t) = - g(t_1 - t)$. This is the case of the  antisymmetric driving,  see dashed and dotted-dashed  lines in Fig.  \ref{fig0}. 

 As a consequence, in the symmetric case (i),  the stationary average velocity  tends to zero when the friction 
 coefficient $\gamma$  tends to zero: $v \to 0$ when $\gamma \to 0$; if $\gamma \ne 0$ then generically
 $v\ne 0$. It is illustrated  in Fig. \ref{figA}  for  $\phi=\pi/2, 3\pi/2$.
In the asymmetric case (ii), the stationary average velocity  tends to zero when the friction 
 coefficient $\gamma$  tends to infinity (the overdamped regime): $v \to 0$ when $\gamma \to \infty$;   
 if $\gamma < \infty$ then generically  $v\ne 0$. Let us note that contrary to the symmetric driving,  for  $\gamma \to 0$ the velocity  $v\ne0$, cf.  Fig. \ref{figA}.  So, it means that the transport is generated by deterministic dynamics. 

We  consider the case of the symmetric driving with $\phi =\pi/2$ for the biharmonicity $\varepsilon = 0.5$ (see Fig. {\ref{fig0})   and study the role of dissipation characterized by the friction coefficient $\gamma$. This is the case when for  $\gamma =0$ the stationary average velocity $v=0$. When the friction coefficient increases starting out  from zero,  the average velocity becomes non-zero as is illustrated in Fig. \ref{figA}.  
The average velocity as a function of $\gamma$ displays non-monotonic dependence exhibiting maxima and minima. 
Moreover, it passes through zero and the current reversal phenomena can be detected.  Because for $\gamma = 0$ 
the velocity $v=0$ and for $\gamma \ne 0$  generically the velocity $v\ne 0$,  this case is called the  
dissipation-induced symmetry breaking \cite{Gommers2005}: the coupling to thermal bath is enough to break the time 
inversion  symmetry.  
We note that for a fixed damping $\gamma$, the average velocity for the phase $\phi=3\pi/2$  takes exactly the opposite sign to the case $\phi =\pi/2$. 

Now, let us consider the antisymmetric case $\phi = 0$. For $\gamma =0$, the velocity $v \ne 0$. The weak dissipation diminishes the stationary velocity in comparison to the dissipationless case. The dependence $v(\gamma)$ is also non-monotonic with minima and maxima. As in the symmetric case, 
the case with the phase $\phi = \pi$ can be obtained from the case $\phi =0$ by the relation 
$v(\phi = \pi) = - v(\phi =0)$.  
\begin{figure}[htbp]
\includegraphics[width=0.9\linewidth]{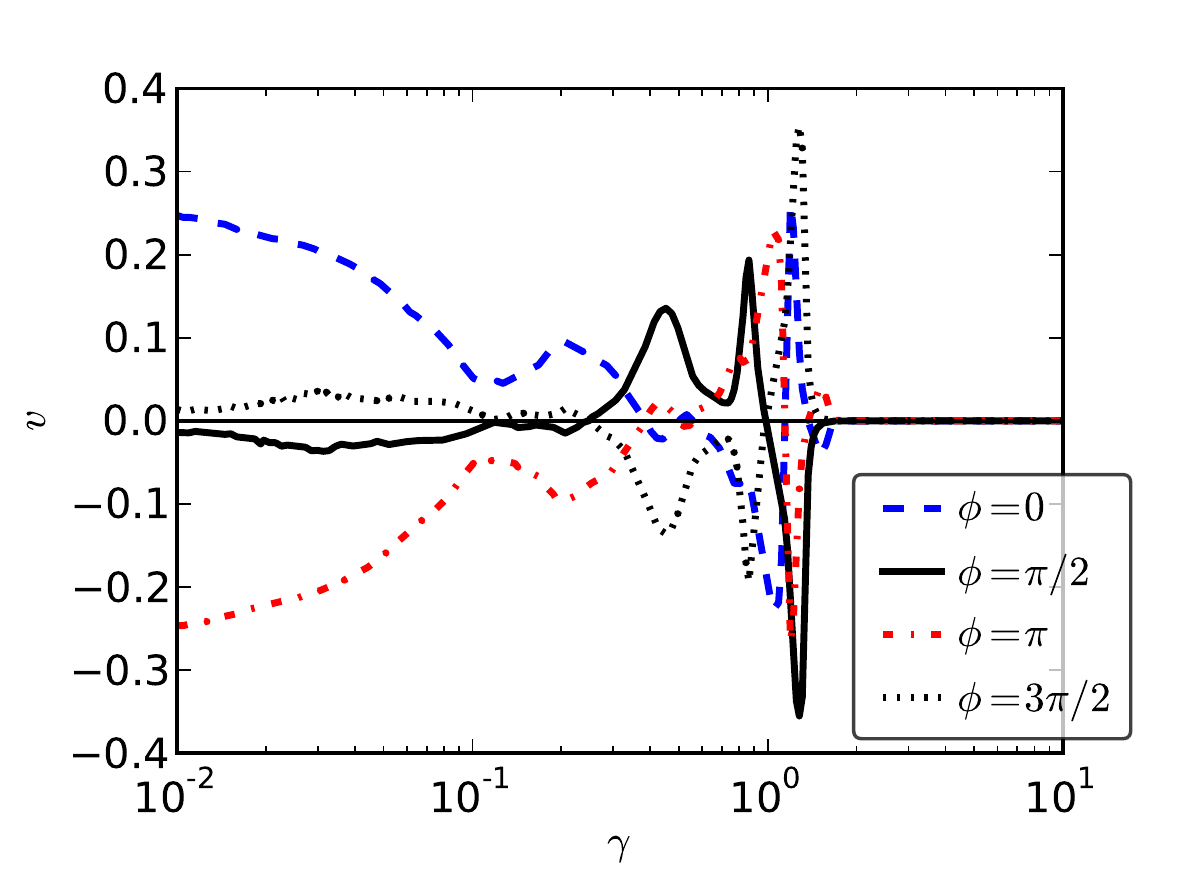} 
\caption{(color online) The stationary average velocity $ v $ as a function of  the friction coefficient 
$\gamma $ is depicted 
for selected values of the relative phase $\phi$. 
For $\phi=\pi/2, 3\pi/2$  the driving is symmetric while for $\phi =0, \pi$ it is antisymmetric.  
Other parameters read: $ a = 4.2$, $ \omega = 4.9$, 
$ \varepsilon = 0.5$ and $ D_0 = 0.001$.
}
\label{figA}
\end{figure}

\begin{figure}[htbp]
\includegraphics[width=0.99\linewidth]{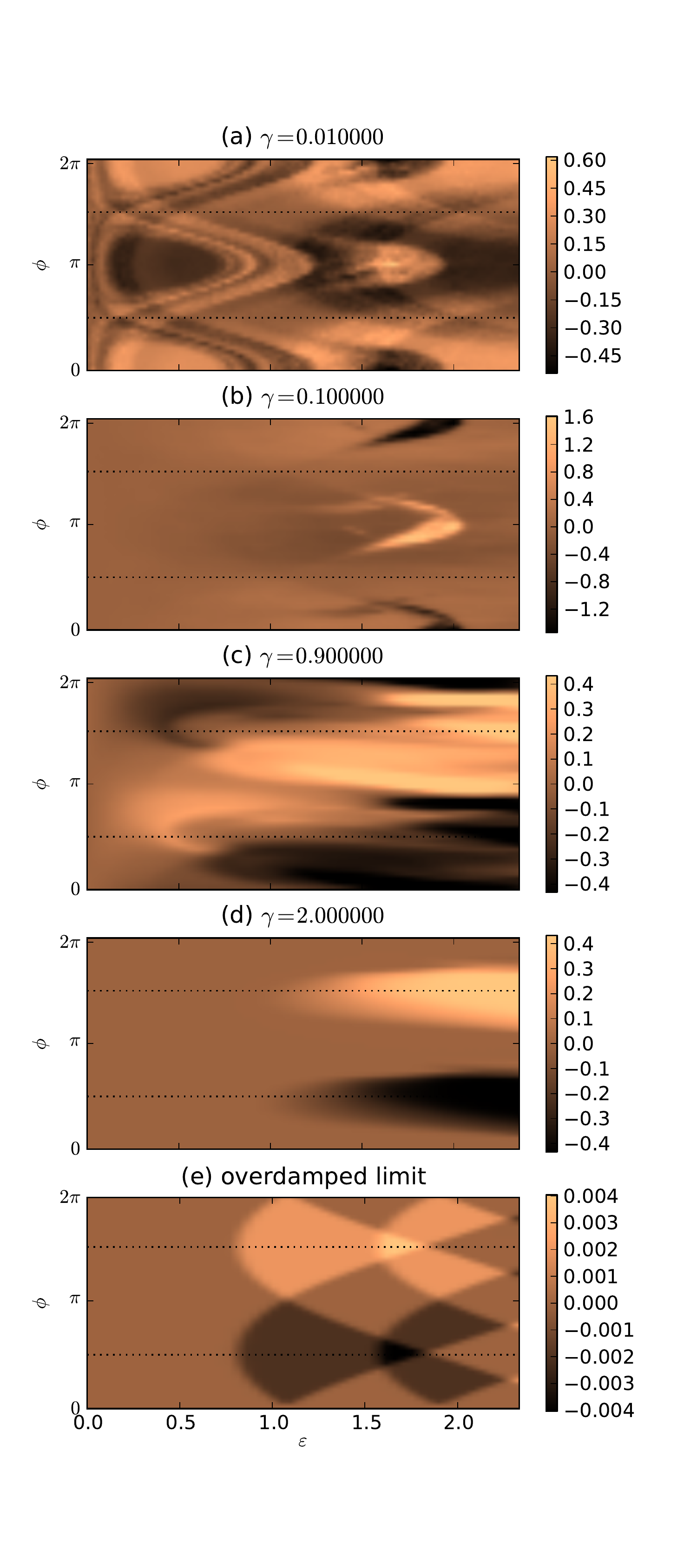} 
\caption{(color online) Influence of the second harmonic of the external force $g(t)$
on  transport properties of the system (\ref{NLbw}).
Dependence of the drift velocity on  both  the relative amplitude $\varepsilon$ (horizontal axis) and the relative  phase $ \phi $  (vertical axis) is depicted for various damping constants 
$ \gamma = 0.01, 0.1, 0.9, 2.0$ and for the overdamped limit (top to bottom). 
Other parameters are: $ a = 4.2 $,
$ D_0 = 0.001 $ and $ \omega = 4.9 $.
Black dotted lines are plotted on all panels showing the phases for which 
the driving force $g(t) $ possesses the reflection symmetry $ t \to -t $,  i.e. for $ \phi = \pi/2$ and $ 3\pi/2 $. 
}
\label{fig3}
\end{figure}

\subsection{Arbitrary shape of driving}

In previous subsection we focused on specific values of the phase. Here we  present 
the numerical investigation of the 3D parameter space $ \{\phi, \varepsilon, \gamma \}$. 
For phases different than just mentioned above, we reveal also asymmetric external  biharmonic  
signals.  
In the Fig. \ref{fig3}, the average velocity is presented in color plots for four different 
damping constants $ \gamma = 0.01, 0.1, 0.9, 2 $ (panels a--d respectively) and additionally 
for the overdamped limit (panel e). On the abscissa we vary the  amplitude $\varepsilon$ of the second component of the signal 
  $ g(t) $ and on ordinate we present phase $ \phi \in [0,2 \pi] $.
Light colors denote positive average velocity.  Color becomes darker for values of $v$ close to zero 
and eventually turn to dark--gray and black for negative valued average velocities.

For the weak friction the average velocity has reflection symmetry 
$ v(\pi +\phi) = v(\pi - \phi) $ as we would expect from a system prepared very close to 
the limit of the frictionless or Hamiltonian systems, because then the relation 
$v \approx \sin(\phi + \pi/2)$  is quite well satisfied \cite{flachEPL}. We plotted black dotted lines on each panel 
to guide the reader to the point where the driving force $ g(t) $ possesses the reflection 
symmetry, i.e., for $ \phi = \pi/2$ and $ 3\pi/2 $.

As we increase the friction coefficient system loses its previous symmetry 
 and becomes  non-symmetric as one can easily see on panels (b) and (c).
In other words -- in the situation where both characteristic times in the system $ \tau_\gamma $ and 
the period  
$ T $  of the driving take more or less  the same value, the battle between periodic stimulation and damping (not strong enough 
to suppress the driving influence quickly with possibility of additional energy cumulation) causes the whole
irregular dynamics as seen on the central panel (c) of  Fig.  \ref{fig3}.
If we, however, analyze situation with strong damping   the picture again gains the symmetry but now  of a different kind,  i.e. 
$ v(\pi +\phi) = - v(\pi - \phi) $, cf. panel (e) in Fig. \ref{fig3}.   The close inspection of all case presented in Fig. \ref{figA} and  Fig. \ref{fig3} leads to the important conclusion that for a fixed set of all parameters, there is the shift-symmetry  of  the stationary velocity  with respect to the phase, i.e., 
\begin{eqnarray}
 \label{shift}
v(\phi) = - v(\phi+\pi). 
\end{eqnarray}
This relation is a particular case of a more general relation 
\begin{eqnarray}
 \label{gen}
v(-\varepsilon) = - v(\varepsilon)
\end{eqnarray}
which  follows from the symmetry considerations.  One can note that the transformation 
$\phi \to \phi + \pi$ is equivalent to the transformation $\varepsilon \to -\varepsilon$. 
The same relation holds  true if, instead  of the second harmonics,  we apply a constant force $F$.  Then of course $v(-F) = - v(F)$ \cite{MacKos2008}.  Remember that for any set of parameters the stationary average velocity 
$v=0$ when $F=0$ or $\varepsilon =0$ 
\begin{figure}[htbp]
\includegraphics[width=0.99\linewidth]{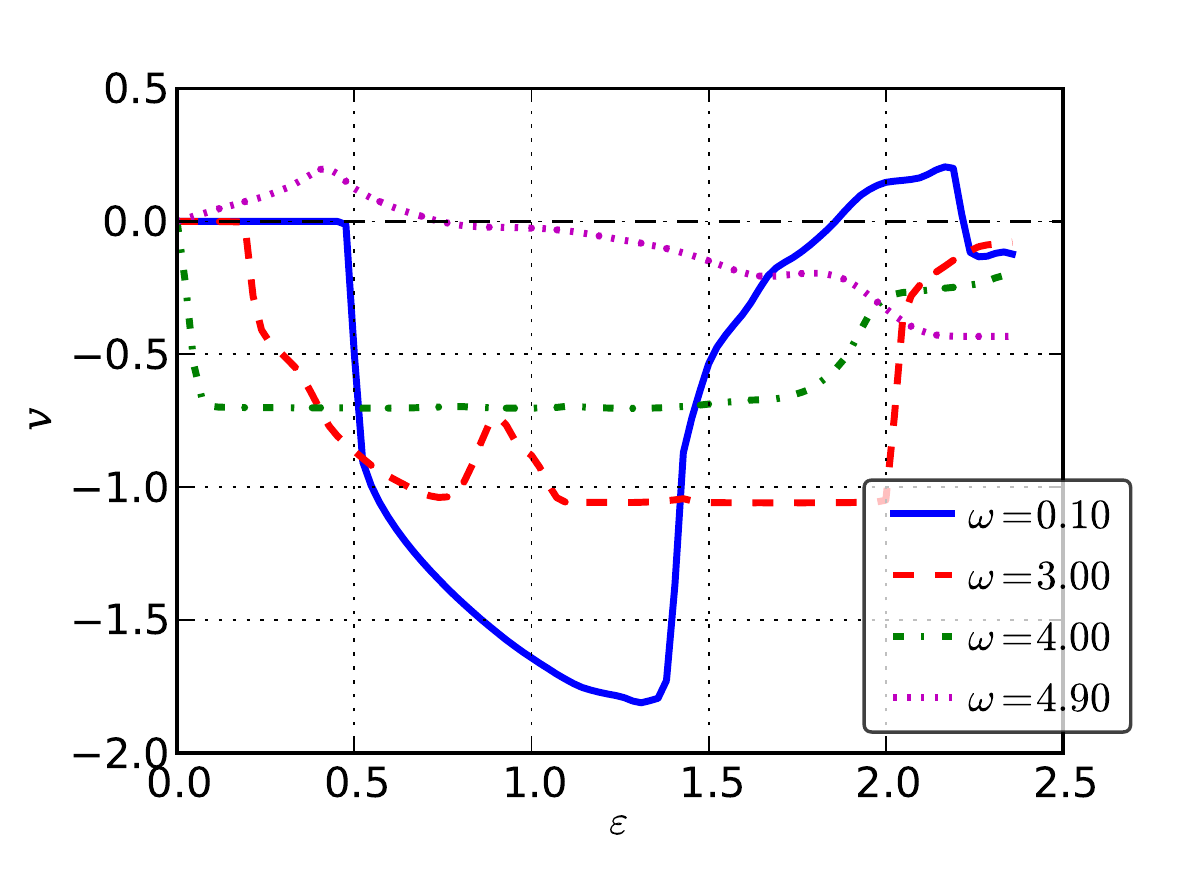} 
\caption{(color online) The stationary averaged velocity {\it vs.} the relative amplitude $\varepsilon$ 
of the second harmonics for four  values of the angular frequency $\omega$ of the signal  g(t).  Other parameters are:  $ a = 4.2 $,  $\gamma = 0.9 $,
$ \phi = \pi/2$ and $ D_0 = 0.001 $.}
\label{fig1}
\end{figure}

\subsection{Controlling transport by  symmetric  signals}

We  analyze  the case when  the external driving is symmetric. We set the phase of the second harmonics to $\phi = \pi/2$ 
(see black solid curves in  Fig.  \ref{fig0}). We check the system 
response to the signal against the relative amplitude of the 
second harmonics $ \varepsilon $ for the range starting from $0$ and 
ending at the value higher than doubled base driving amplitude $a$. In 
Fig.  \ref{fig1} these characteristics are plotted for  selected 
 driving frequencies $ \omega = 0.1, 3, 4, 4.9$. 
From numerical analysis it follows that the average velocity changes its sign by
varying the parameter $ \varepsilon $ for all inspected frequencies of
the external driving. It means that the shape of the external signal
can  control   values and direction of the net velocity in the
system. The current reversal can be multiple ($\omega=0.1$), akin to
the situation described in \cite{KosturMCR}. Keeping $\varepsilon$
constant at a certain level usually the direction of the average
motion of Brownian particles changes its sign for the different
values of the driving frequency ($ \varepsilon = 0.5$
or $ 2.0 $). On the contrary there are regimes within the scanned
parameter space $ \{\varepsilon, \omega\} $ where regardless the values
of $\omega$ chosen the system response is qualitatively the same.

%
\begin{figure}[htbp]
\includegraphics[width=0.99\linewidth]{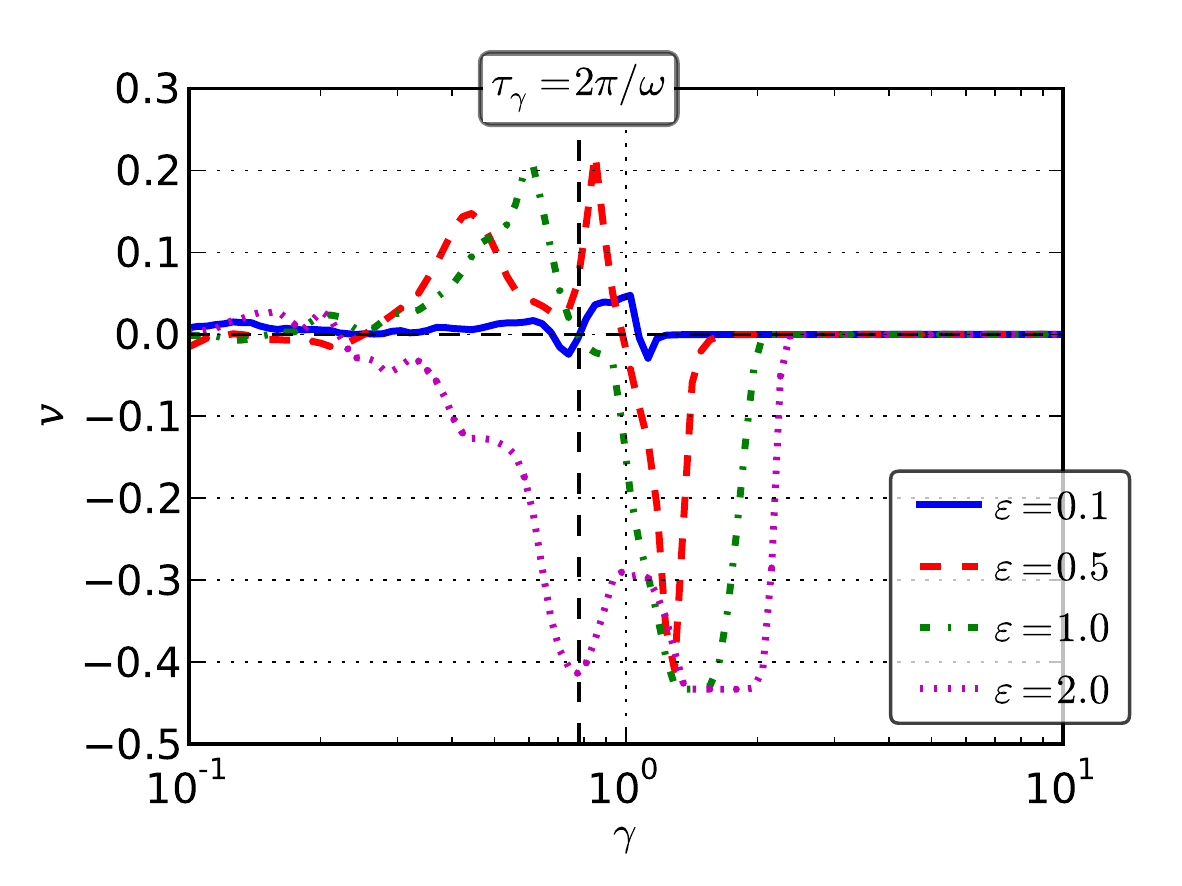} 
\caption{(color online) Logarithmic dependence of the average velocity on the friction
coefficient $\gamma$  is plotted for four relative amplitudes $\varepsilon$ of the second harmonics 
 of the external driving  $g(t)$. 
Vertical black dashed line marks the point of the critical value of the friction
coefficient, for which two characteristic times, relaxation time of the velocity  $\tau_{\gamma}=1/\gamma$ and 
period $T=2\pi/\omega$ of the driving force, are equal. One can easily notice rich behavior of 
the average velocity around this specific value.
Other parameters are: $ a = 4.2 $, $ \omega = 4.9 $,
$ \phi = \pi/2 $ and $ D_0 = 0.001 $.
}
\label{fig2}
\end{figure}

Next we explore the transport properties for the Brownian particle moving in the viscous
environment with different friction coefficients. We examine the character of the system 
response against the signal of the different shape which we can control by tuning the 
parameter $ \varepsilon $ (see Fig.  \ref{fig0} for details).
There are two alternative limits for the viscous system behavior - Hamiltonian where
system is frictionless \cite{hr001,hr002,hr003} 
and overdamped where the characteristic relaxation time for the
velocity $ \tau_\gamma = 1 / \gamma $ is very long. 
Between those two peripheries there is a region of moderate damping which seems to be
the most intriguing \cite{MacKos2007,MacKos2008,MacKos2005,RatchetANM}. It provides rich spectrum 
of the very interesting phenomena and therefore we are going to focus on this particular 
domain in the following. 

In  Fig. \ref{fig2} the reflection of the impact of different shapes 
and strengths of external driving for the friction constant $ \gamma $ in the range
from 0.1 to 10 can be found. This means that the characteristic relaxation time passes from
$ 10 $ to $ 0.1 $. If we refer this time to the second characteristic time of importance
for system (\ref{NLbw}), namely the period of the external driving 
$ T = 2 \pi / \omega  \simeq 1.28$, one can see that the point where
both characteristic times are of the same order can be identified more or less in the middle
of the chosen region of analyzed damping constants.
Indeed, after examining of Fig. \ref{fig2}, one can easily reveal most exciting features around
essential value of the damping constant marked by the vertical dashed black line on the plot.
At low  friction, the  average velocity is close to zero. When we, however, increase the 
friction coefficient to the value of around $ \gamma = 0.3 $ the  system starts to react in 
a different way depending on the relative strength of the second harmonic $ \varepsilon $. For 
strengths less then or equal to 1 the current becomes positive, while for $ \varepsilon = 2 $ system
reacts with the opposite sign. This gives a possibility to control the transport simply by varying 
the strength of the second source of the external field. When we go even further and arrive to  the vicinity
of the critical point  $ \tau_\gamma = 2 \pi / \omega$, the previous positive valued current starts to
drop, crosses zero and becomes negative quite steeply. Surprisingly values of average velocities for 
all strengths higher than 0.5 possess almost the same negative values just above $ \gamma = 1 $.
Additional enlargement of the friction leads to reduce of the transport possibilities of the 
system. It does not reach zero, but decreases of several orders of magnitude -- see panel (e) on
Fig. \ref{fig3} for details. By setting the strength to zero we end up with the antisymmetric 
force and with zero current for any value of the friction constant due to the symmetry reasons.

\section{Summary}

We have explored transport properties  of the Brownian 
particles in a symmetric potential, driven by the time periodic  biharmonic 
signals.  We have demonstrated how the symmetry of driving force  influences   
the  transport features. There exists two limits: overdamped and 
frictionless. It turns out that in those two limits different types of 
symmetry  exclude transport. In the frictionless case the system 
is time-reversible, thus the symmetric driving cannot distinct the 
direction. On the other hand, in the case of overdamped motion the 
antisymmetric driving leads to zero current. In all other 
cases, as the Curie principle suggests, the particle has generally non-zero 
average velocity. The closer inspection shows that the magnitude and 
sign of the current has complex structure in the parameter 
space. Typically, the multiple current reversals occur, when one of 
the system parameters is changed. 

In this paper,  thermal noise  $\xi(t)$ in Eq. (\ref{Lan}) is assumed  to
be  white noise of  zero  correlation time.  In real systems the correlation time of thermal
fluctuations is never zero.  In many situations  this approximation is very well but 
there are also situations where  the white-noise
approximation fails and a different treatment based e.g. on the generalized Langevin equation
should  be used  \cite{PREANM}. However,  it  is  essentially beyond the scope of the paper and requires separate  investigations. 

Finally, let us remind that the Langevin equation {(\ref{Lan})  has similar form  
as an equation of motion for the phase
difference $\Psi=\Psi(t)$  between the macroscopic wave functions of the
Cooper pairs on both sides of the Josephson junction. 
The quasi-classical dynamics of the resistively and capacitively
shunted Josephson junction, which is well known in the literature as  the Stewart-McCumber
model \cite{stewart,mccumber,kautz},   is described by the following equation
\begin{eqnarray} \label{JJ1}
\Big( \frac{\hbar}{2e} \Big)^2 C\:\ddot{\Psi} + \Big( \frac{\hbar}{2e} \Big)^2 \frac{1}{R} \dot{\Psi}
+ \frac{\hbar}{2e} I_0 \sin \Psi \nonumber\\ = \frac{\hbar}{2e} I(t) + \frac{\hbar}{2e}
\sqrt{\frac{2 k_B T}{R}} \:\xi (t) .
\end{eqnarray}
The left hand side  contains three additive
current contributions: a displacement current due to
the capacitance $C$ of the junction, a normal (Ohmic) current characterized
by the normal state resistance $R$ and a Cooper pair tunnel current characterized by
the critical current $I_0$.   In the right hand side, $I(t)$ is an external current. 
Thermal fluctuations of the current are taken into
account according to the fluctuation-dissipation theorem and satisfy the Nyquist
formula associated with the resistance $R$. 
It is an evident correspondence between two models:  the coordinate $x=\Psi -\pi/2$,    the mass $m=(\hbar /2e)^2 C$,   the friction coefficient $\gamma =
(\hbar/2 e)^2(1/R)$, the barrier height $\Delta V = (\hbar/2 e) I_0$ and the
period $L=2\pi$.  
 The biharmonic signal $G(t)$ in Eq. (\ref{biforce}) corresponds to the external  current $I(t)$ . 
The velocity $v=\dot x$ corresponds to  the voltage $V$ across the junction. 
So, all transport properties can be tested in the setup consisting of a resistively and
  capacitively shunted Josephson junction device.

\section*{Acknowledgment} 

The work supported in part by the MNiSW Grant N202 203534 and the Foundation for Polish Science 
(L.  M.). The authors  thank M. Januszewski for preparing the  precise program  
(http://gitorious.org/sdepy)  that we have used for  numerical calculations.
We would like to acknowledge Peter H\"anggi, our friend and  mentor, for long-term  collaboration, inspiring,  motivating and never ending - not only scientific - discussions.

\end{document}